\documentclass[aps, prd, preprint, nofootinbib, superscriptaddress]{revtex4-2}
\usepackage{amsmath, amssymb, graphicx, hyperref}
\usepackage{caption, subcaption}
\usepackage[english]{babel}
\def\bea{\begin{eqnarray}} \def\eea{\end{eqnarray}} \def\ba{\begin{array}}
\def\be{\begin{equation}} \def\ee{\end{equation}} 
\def\bi{\begin{itemize}} \def\ei{\end{itemize}} 
\def\ea{\end{array}} \def\ben{\begin{enumerate}} \def\een{\end{enumerate}}

\newcommand{\eqn}[1]{(\ref{#1})}

\def\br{\nonumber\\}

\def\g{\gamma}

\begin{document}
\title{Holographic entanglement entropy for $Lif_4^{(2)}\times {S}^1\times S^5$\\ spacetime with string excitations}
\author{Sabyasachi Maulik}
\email{sabyasachi.maulik@saha.ac.in}
\author{Harvendra Singh}
\email{h.singh@saha.ac.in}
\affiliation{Theory Division, Saha Institute of Nuclear Physics, 1/AF, Bidhannagar, West Bengal 700064, India}
\affiliation{Homi Bhabha National Institute, Anushakti Nagar, Mumbai 400094, India
\vspace{1.5 em}}
\begin{abstract}
The (F1,D2,D8) brane configuration with $Lif_4^{(2)}\times {S}^1\times S^5$ geometry is a known Lifshitz vacua supported by massive $B_{\mu\nu}$ field in type IIA theory. This system  allows exact IR excitations which couple to massless modes of the fundamental string. 
Due to these massless modes the solutions have a flow to a dilatonic $Lif_4^{(3)}\times S^1\times S^5$ vacua in IR. We study the entanglement entropy on the boundary of this spacetime for the strip and the disc subsystems. To our surprise net entropy density of the excitations at first order is found to be independent of the typical size of subsystems. We interpret our results in the light of first law of entanglement thermodynamics.
\end{abstract}
\maketitle
\vfill
\section{Introduction}
The  gauge-gravity correspondence
 \cite{Maldacena:1997re, Gubser:1998bc, Witten:1998qj}
has got a nonrelativistic version  where
 strongly coupled quantum theories at critical points
 can be studied  \cite{Son:2008ye, Balasubramanian:2008dm, Herzog:2008wg, Hartnoll:2008vx, Hartnoll:2008kx, Maldacena:2008wh, Denef:2009tp, Kachru:2008yh, Singh:2009tq, Singh:2010rt, Balasubramanian:2010uk, Singh:2010zs, Singh:2012un, Narayan:2012hk, Singh:2013iba, Narayan:2012ks, Singh:2017wei, Singh:2018ibp, Mishra:2018tzj, Taylor:2015glc}.
Some of these quantum systems  involve  strongly coupled fermions 
 at finite density or it may simply be a gas of ultra-cold atoms 
\cite{Son:2008ye, Balasubramanian:2008dm}. 
In the studies involving 
`nonrelativistic' Schr\"odinger spacetimes the 4-dimensional
  spacetime geometry generally requires supporting
 Higgs like field such as  massive vector field 
 \cite{Herzog:2008wg, Denef:2009tp, Son:2008ye} or a tensor field. 
The  spacetimes possessing a Lifshitz symmetry
 provide  similar holographic dual description of  nonrelativistic 
quantum theories living on their  boundaries \cite{Kachru:2008yh}, see \cite{Taylor:2015glc} for a review.  

In this work we shall mainly study entanglement entropy of the 
excitations in asymptotically 
$Lif_4^{(a=2)}\times S^1\times S^5$ background. The latter is 
a Lifshitz vacua in massive type IIA (mIIA) theory \cite{Singh:2017wei, Singh:2018ibp} with dynamical exponent of time being $a=2$. The massive type IIA theory \cite{ROMANS1986374} is a ten-dimensional maximal supergravity where the antisymmetric 
tensor field is explicitly massive. The theory also includes a positive cosmological constant related to mass parameter. Due to this structure the mIIA theory provides a unique setup to study Lifshitz   solutions. Particularly the  $Lif_4^{(2)}\times S^1\times S^5$
solution is a background generated by the bound state of $(F1,D2,D8)$ branes \cite{Singh:2017wei}
\bea\label{sol2a9n}
&&ds^2= L^2\left(- {dt^2\over  z^4} +{dx_1^2+dx_2^2\over z^2}+{dz^2\over
z^2}  +{dy^2\over  q^2} + d\Omega_5^2 \right) ,\br
&&e^\phi=g_0 ,
 ~~~~~C_{(3)}= -{1 \over g_0}
{L^3\over  z^4}dt\wedge dx_1\wedge dx_2, \br &&
B_{(2)}=  { L^2\over q  z^2}dt\wedge dy  
\eea 
The metric and the form fields have explicit invariance under constant scalings (dilatation); 
$z\to \lambda z,~t\to \lambda^2 t,~
x_i\to \lambda x_i, ~y\to y$. The dynamical exponent of time is $2$ here.
The background describes a strongly coupled nonrelativistic 
quantum theory at the UV critical point. 
\footnote{Analogous T-dual  solution do also exist in type IIB  theory with 
constant axion flux  switched 
on \cite{Balasubramanian:2010uk}}  

It is worthwhile to study excitations of the
$Lif_4^{(2)}\times S^1\times S^5$
 vacua as it immediately provides us a prototype  
$Lif_4^{(2)}$ background in four dimensions which is holographic dual to 3-dimensional Lifshitz theory on its boundary. The  excitations would tell us how this Lifshitz theory behaves
near its critical point. Particularly we shall study a class of string like excitations which themselves form solutions of massive IIA sugra 
and explicitly involve $B$-field \cite{Singh:2018ibp}. 
These also induce running of  dilaton as well. It is observed that the resulting RG flow in the deep IR can be described simply by ordinary type IIA  theory. The reason for this is due to the fact that the contributions of massive stringy modes decouple from the low energy  dynamics of the theory in the IR, far away from UV critical point \cite{Singh:2018ibp}. 

In this report we aim to study holographic entanglement entropy (HEE) \cite{Ryu:2006bv, *Ryu:2006ef, *Hubeny:2007xt} of the excited Lifshitz subsystems which are either a disc or a strip in a perturbative framework. A critical observation is that for small sized systems the entanglement entropy density remains constant at first order. That is, 
the first order contributions to the entropy density remain independent of the
size ($\ell$) of the subsystem. This is a peculiarity and quite unlike  relativistic CFTs where usually the entropy density (of excitations) is linearly proportional to the typical size of the subsystem \cite{Bhattacharya:2012mi}. 
We discover that the resolution lies in the nature
of the chemical potential ($\mu_E$) for the Lifshitz  system. We gather  evidence that 
suggests that  energy density (of excitations) falls off with the size of system as $\propto 1/\ell^2$. Furthermore the $1/\ell^{2}$ dependence is exactly same as the entanglement temperature behaviour in the Lifshitz theory. 
Notwithstanding these peculiarities,  
 the entropy of excitations consistently 
follows the first law of entanglement
thermodynamics \cite{Bhattacharya:2012mi,Allahbakhshi:2013rda} up to first order.
\par In addition, we also carry out a calculation of entanglement entropy at second order for both disc and strip subsystems. Contributions arising at this order bestow an explicit $\ell$ dependence upon the entropy. We argue how the first law can still be obeyed by modifying our chemical potential $(\mu_E)$ and entanglement temperature $(T_E)$. A similar argument was put forward in \cite{Mishra:2015cpa} for asymptotically AdS spacetime.

The unusual symmetry of Lifshitz spacetime makes it a good background to study novel features of entanglement in a non-relativistic quantum theory at zero temperature \cite{Son:2008ye, Balasubramanian:2008dm, Kachru:2008yh}. It is well known that for such systems, e.g. a particle in a one-dimensional box the momentum of the particle scales with the length as $ p \propto \frac{1}{\ell}$ and the energy $\mathcal{E} \propto \frac{1}{\ell^2}$; our calculations of entanglement entropy also support this explicit size dependence of energy, as shown in equation \eqref{enr34}. We hope our work will help shed some light on holographic treatment of non-relativistic quantum systems at strong coupling that are often interesting in e.g. condensed matter theory.

\par The rest of the paper is organized as follows: in section \ref{sec2} we review salient features of $Lif^{(2)}_4\times S^1\times S^5$ vacua with IR excitations in mIIa theory. The holographic entanglement entropy for a disc subsystem is calculated in section \ref{sec3}. In section \ref{sec4} we carry out similar analysis for strip subsystem at first and second orders, section \ref{sec5} contains the conclusion.

\section{$Lif^{(2)}_4\times S^1\times S^5$ vacua and excitations}\label{sec2}
 
 The massive type IIA supergravity theory is the only known maximal 
supergravity in ten dimensions which allows  massive string $B_{\mu\nu}$  
field and a mass dependent cosmological constant \cite{ROMANS1986374}.
The cosmological constant 
 generates a  nontrivial  potential term  for the dilaton 
field. The mIIA theory  does not admit  flat  Minkowski 
 solutions. 
Nonetheless the theory gives rise to  well known  
Freund-Rubin type vacua $AdS_4\times S^6$   \cite{ROMANS1986374},  
the  supersymmetric domain-walls 
or D8-branes \cite{Polchinski:1995mt, Bergshoeff:1996ui,Witten:2000mf, Hull:1998vy, Haack:2001iz}, 
  $(D6,D8)$, $(D4,D6,D8)$ 
bound states \cite{Singh:2001gt, Singh:2002eu} and Galilean-AdS geometries \cite{Singh:2009tq, Singh:2010rt}. 
 In all of these massive tensor field
plays a key role.
Under the `massive' T-duality \cite{Bergshoeff:1996ui} the D8-branes  
 can be mapped  over to the axionic D7-branes of type IIB string theory
and vice-versa.  
The  $B$-field also plays  important role in obtaining  
non-relativistic Lifshitz solutions \cite{Singh:2017wei, Singh:2018ibp}. 
The latter solutions are of no surprise in mIIA theory,
as an  observed  feature in four-dimensional AdS gravity theories
has been that in order to obtain  non-relativistic 
solutions one needs to include 
massive (Proca)  gauge fields in the gravity theory \cite{Son:2008ye}. 
Other  different situations where massless vector fields 
can give rise to  non-relativistic vacua, 
  involve  boosted black D$p$-branes
compactified along lightcone direction \cite{Singh:2010zs, Singh:2012un}. These latter class of solutions are also called 
hyperscaling (or conformally) Lifshitz  vacua \cite{Narayan:2012hk}.   

Particularly the $a=2$ Lifshitz vacua with IR excitations in mIIA theory
can be written as \cite{Singh:2018ibp} 
\bea\label{sol2a9}
&&ds^2= L^2\left(- {dt^2\over  z^4h} +{dx_1^2+dx_2^2\over z^2}+{dz^2\over
z^2}  +{dy^2\over  q^2h} + d\Omega_5^2 \right) ,\br
&&e^\phi=g_0 h^{-1/2},
 ~~~~~C_{(3)}= -{1 \over g_0}
{L^3\over  z^4}dt\wedge dx_1\wedge dx_2, \br &&
B_{(2)}=  { L^2\over q  z^2}h^{-1}dt\wedge dy  \ ,
\eea 
where the harmonic function $h(z)= 1+{z^2\over z_{I}^2}. $
The parameter $z_I$ is related to the charge of the NS-NS 
strings. 
The excitations  involve $g_{tt}$ and $g_{yy}$
 metric components, and leaving the  $x_1,x_2$ 
plane (worldvolume directions of D2-branes) 
unaffected.\footnote{
Here  $L={2\over g_0  m l_s}$, 
and $m$ being the mass parameter in the mIIA action. (We would set $l_s=1$ and
 $g_0=1$.) 
 The  constant $q$ is a free (length)  parameter 
and $g_0$ is weak string coupling.
Note $L$ is dimensionless parameter, it
determines overall radius of curvature of the spacetime.
 Therefore Romans' theory with 
  $m \ll{ 2\over g_0 l_s}$ would be preferred here 
so that  $L\gg 1$ in  the solutions \eqn{sol2a9},  else
  these classical vacua cannot be trusted. Also, from  the D8
brane/domain-wall correspondence  in \cite{Bergshoeff:1996ui}, one typically
expects $m \approx {g_0 N_{D8} \over l_s}$, a value which is definitely
well within ${ 2\over g_0 l_s}$ for a finite number of $D8$ branes, $N_{D8}$, 
in these backgrounds. }
 The  excitations do also  induce a running of dilaton field. 
The $B_{ty}$ component of the string field  
 is also coupled to the  excitations. Since
$h\sim 1$ as  $z\to 0$, 
these excitations form normalizable modes ($z_I$ would correspond to 
adding relevant operators in the boundary Lifshitz theory). 
The solution \eqn{sol2a9} asymptotically flows to  weakly coupled  
  regime  in the UV (note that the string coupling,\ $g_0<1$). 
While, in the
deep IR region, with $z\gg z_I$ where $h\approx {z^2\over z_I^2}$,
the   vacua is  driven to another  
weakly coupled Lifshitz regime.  For $z\gg z_I$,  
the  IR geometry transforms  to dilatonic
  $Lif_4^{(3)}\times S^1\times S^5$ solution.
This solution enables us to study  the effect 
of the excitations in 
 $a=2$  Lifshitz theory. Note the $z_I$ dependent excitations 
at zero temperature
are mainly in the form of charge excitations, along with nontrivial 
entanglement chemical potential, as we would see next.

\section{ Entanglement  of a disc subsystem}\label{sec3}
 
For asymptotically AdS space-time dual to a CFT, the entanglement entropy can be calculated by the Ryu-Takayanagi formula \cite{Ryu:2006bv, *Ryu:2006ef}. We assume the same is true for an asymptotically Lifshitz space-time, dual to a non-relativistic field theory with Lifshitz scaling symmetry. We consider a round disc of radius $\ell$ at the center of the $x_1,x_2$ plane with its boundary identified with the corresponding boundary of  $2d$  Ryu-Takayanagi surface lying inside the Lifshitz bulk geometry \eqn{sol2a9}. We shall assume $y$ is a compactified direction 
\begin{equation}
	y\sim y +2\pi r_y\;.
\end{equation}
In  radial coordinates $(r=\sqrt{x_1^2+x_2^2})$ 
the Ryu-Takayanagi area functional \cite{Ryu:2006bv, Ryu:2006ef} 
for static bulk surface is given by 
\bea \label{areaint1}
{\cal A}_\gamma = 2\pi L^2 \int^{z_\ast}_{\epsilon} dz 
 {r\sqrt{1+r'^2}\over z^2} h^{1\over2}\,,
\eea
where, $r'={dr\over dz},~h(z)=\left(1+ {z^2\over z_I^2 }\right)$ and 
$\epsilon \ll \ell$ is UV cut-off  
of the Lifshitz theory. We need to extremize the area integral by solving the Euler-Lagrange equation for $r(z)$
\begin{multline}
	2zrr''h(z) - 4rr'^3h(z) - 4rr'h(z) - 2zr'^2h(z) -2zh(z) - zrr'^3h'(z) - zrr'h'(z) = 0\,,
\end{multline}
It is impossible to analytically calculate the full area integral \eqref{areaint1}. To facilitate our job, therefore, we restrict ourselves to small subsystems, with  $\ell\ll z_I$. In this domain, we can make a perturbative expansion and obtain solutions order by order in the dimensionless ratio ${\ell\over z_I}$; such that $r(z)=r_{(0)}+ r_{(1)}+ \cdots$, and correspondingly we would write $${\cal A}_\g ={\cal A}_0 +{\cal A}_1 +\cdots\;,$$ for small  $\ell$. Our immediate interest is in calculating terms up to leading order and 
first order only in the ${\ell\over z_I} $ expansion. 
\par The equation at zeroth order is
\begin{equation}
	zr_{(0)}r_{(0)}^{\prime \prime} - 2r_{(0)}r_{(0)}^{\prime 3} - 2r_{(0)}r_{(0)}^{\prime} - zr_{(0)}^{\prime 2} - z = 0\,,
\end{equation}
for which  $r_{(0)}=\sqrt{\ell^2-z^2}$ defines the extremal surface (half circle) \cite{Ryu:2006ef, Blanco:2013joa} with the boundary conditions $r_{(0)}(0)=\ell$, and $r_{(0)}(z_\ast)=0$, where $z=z_*$ is the point of return that lies at $z_\ast= \ell$. One then finds that the area

\begin{align}
{\cal A}_0 &= 2\pi L^2 \int^{z_\ast}_{\epsilon} dz \frac{r_{(0)}\sqrt{1 + r_{(0)}^{\prime 2}}}{z^2},\nonumber \\ 
&= 2\pi L^2 \left(\frac{\ell}{\epsilon} - 1 \right).
\end{align}

${\cal A}_0$ being a ground state contribution it obviously remains independent of the parameter $z_I$ of the bulk geometry. This only means that there is no effect of excitations on the leading term. As explained in \cite{Blanco:2013joa}, the first order contribution can be evaluated using only the tree level embedding function and is given by
\bea \label{area1}
{\cal A}_1 &=& 
2\pi L^2 \int^{z_\ast}_{\epsilon} dz r_{(0)} 
{\sqrt{1+r_{(0)}'^2}\over 2\,z_I^2},\br
&=& \pi L^2  \left({\ell^2\over z_I^2}\right).
\eea
From here the complete expression of entanglement entropy of 
a  disc shaped subsystem up to first
order becomes
\bea\label{gb12}
S_E^{Disc}[\ell,z_I]&\equiv& {{\cal A}_\gamma \over 4 G_{4}},\br
&=& S_{E}^{(0)}+
 {\pi L^2 \over  4\,G_{4}}  \left( 
   {\ell^2\over z_I^2} \right),
\eea
where the Newton's constant in $4$D and $5$D are related to the 10-dimensional Newton's constant by $\frac{1}{G_{4}} = \frac{L\,2\pi r_y}{G_5}$ and $\frac{1}{G_5}\equiv {L^5 Vol(S^5)\over G_{10}}$. We shall be using $G_4$ and $G_5$ back and forth in our calculation.\\
The ground state entropy contribution is
\bea
S_{E}^{(0)}= {\pi L^2 \over  2G_{4}}  \left( 
{\ell\over \epsilon} -1\right)\ .
\eea
The equation \eqref{gb12} is a  meaningful expression for entanglement entropy only if we maintain   $\ell\ll z_I$. The first order term explicitly depends on $z_I$, so  small fluctuations of the bulk quantities, like $\delta z_I$, would result in corresponding change in entropy. For a fixed size $\ell$, one could express these variations of the entropy density as
\bea \label{ent34}
{\delta s_E^{Disc}}
={\delta S_E^{Disc}\over 
\pi \ell^2}
=  {L^2 \over 4 G_4} 
   \delta \left( {1\over  z_I^2} \right),
\eea
where $\pi \ell^2$ is the disc area.
Equation \eqref{ent34} provides a complete expression  up to  first order. 
At second order the entropy
will receive  new $z_I$ dependent contributions.

Next, we note that
the right hand side of equation \eqref{ent34} is actually independent of the disc size $\ell$!
On first hand observation this appears very surprising because, according to the first law of 
entanglement thermodynamics \cite{Bhattacharya:2012mi}, we expected that the entropy density of excitations would  have had  $\ell^2$ dependence, namely in the form of inverse temperature (usually entanglement
temperature goes as $T_E^{-1}\propto \ell^a$; and the dynamical exponent of time in our Lifshitz background  is $a=2$). Especially this aspect of the  first law has been found to remain true  in a variety of relativistic CFTs, where entanglement temperature is given by $T_E \propto {1 \over \pi  \ell}$; see for example \cite{Bhattacharya:2012mi, Allahbakhshi:2013rda, Mishra:2015cpa, Mishra:2016yor, Mishra:2018tzj, Ghosh:2017ygi, Bhattacharya:2019zkb}. What, then, is so different for the Lifshitz system described by equation \eqn{ent34}? To understand this phenomenon we first need to get an estimate of the energy associated with the excitations in our system.

\subsection {Energy, winding charge and chemical potential}
 
We now turn to find the energy of excitations of the `massive strings' due to which we have a configuration in equation \eqn{sol2a9},
where we can express $B_{ty}\simeq B_{ty}^{massive}
+B_{ty}^{excitation}$.
Note that we are treating $y$ as a compact direction.
The Scherk-Schwarz compactification \cite{Scherk:1978ta, *Scherk:1979zr, Lavrinenko:1996mp} of the Lifshitz background \eqn{sol2a9} on a circle 
along $y$ gives rise to the following 1-form potential
\be
A_{(1)}={L^2\over q z^2}\left(1+ {z^2\over z_I^2}\right)^{-1} dt.
\ee
It represents a gauge field in the lower dimensional supergravity whose only non-zero component is $A_t$. It can be determined from here that due to  string excitations the net change in the $U(1)$ charge (due to winding strings) is
\be\label{rho12}
 \bigtriangleup \rho 
= {N\over V_2}= {\bigtriangleup Q\over 2\pi r_y V_2}=
{2 L\over  G_5 z_I^2},\ee
where $V_2$ is the area element of $x_1,x_2$ plane, see a calculation in the appendix. The entanglement chemical potential, with the prescription in \cite{Mishra:2015cpa}, 
can be obtained by measuring gauge  field 
at the turning point, 
 namely
\bea\label{chempotdef}
\mu_E \equiv A_t|_{z=z_\ast} = {L^2r_y \over q z_\ast^2}+ \cdots \,,
\eea  
where ellipses denote  sub-leading terms which are not required at first order. This is a logical guess inspired by black hole thermodynamics, where the value of the one form at the black hole horizon is known to give the chemical potential conjugate to the U(1) charge. Even for backgrounds with non-relativistic conformal symmetry as considered in \cite{Maldacena:2008wh}, the Kaluza-Klein gauge field measured at the horizon produces the correct thermal chemical potential. There's no horizon in our bulk space-time; instead, we use the critical point $z_*$ associated with the entanglement wedge.

At leading order we have $z_\ast \simeq \ell$, hence essentially this thermodynamic variable gets uniquely fixed by the 
Lifshitz ground state \eqn{sol2a9n}. 
So for small  $\ell ~ (> 0)$ the chemical potential remains quite important, and we obtain
\bea 
\mu_E\cdot \bigtriangleup \rho \simeq 
{L^2 \over \pi G_4}{1\over z_I^2\ell^2}. 
\eea
 There are no other excitations except the winding strings, the energy density due to
 the excitations can be estimated to be
\begin{equation} \label{enr34}
\bigtriangleup {\cal E}={\cal E}-{\cal E}_0 \simeq 
\frac{1}{2}\mu_E \bigtriangleup \rho = \frac{L^3 r_y}{q G_5} \frac{1}{z_I^2 \ell^2} = {L^2 \over 2\pi G_4}\frac{1}{z_I^2 \ell^2}\,,
\end{equation}
 where ${\cal E}_0$ is the (normalized) energy of the
ground state of our Lifshitz theory\footnote{We do notice an explicit dependence of energy density on the system size; which is unlike relativistic CFT but is a familiar feature in non-relativistic theories, the particle in a box being an immediate example.}. This is the only meaningful deduction we can make from here, particularly in absence of a direct method to evaluate full stress-energy tensor of the Lifshitz theory.\footnote{ There is an early work \cite{Ross:2009ar} but it does not include dilatonic scalar field excitations like in our background. In contrast in asymptotically AdS spacetimes one knows how to obtain stress-energy tensor by doing Fefferman-Graham expansion near  AdS boundary \cite{Balasubramanian:1999re,*Kraus:1999di,*Bianchi:2001kw}. Perhaps something similar could also be done in the Lifshitz case involving dilaton field.} Assuming that the entanglement temperature of the 3-dimensional
$a=2$ Lifshitz system  faithfully behaves as \cite{Bhattacharya:2012mi}
 \be\label{spheretemp1}
T_E = {4\over \pi \ell^2}\,,\ee
we determine that the ratio 
$${\mu_E \over T_E} = {\pi L^2 r_y\over 4  q}\,,$$ 
 is indeed independent of  $\ell$. 
Essentially this  ratio seems to get uniquely fixed by the Lifshitz ground state \eqn{sol2a9n} at the leading order. Note that the excitations seem to have no effect on it. The analysis also implies that the energy density and the entanglement temperature both fall off with the system size $\ell$ at the same rate, and the ratio
\bea
{\bigtriangleup {\cal E} \over T_E}= 
{ \pi L^3  r_y \over 4 q G_5 z_I^2} 
\equiv {1\over 2} {k_E N\over  V_2}, 
\eea
stays fixed for small discs. However this ratio does depend on the excitations 
namely through $z_I$. 
In the second equality we have preferred to view dimensionless quantity
 $k_E={\pi  L^2 r_y\over 8 q}$ as being 
analogous to the Boltzmann constant in usual 
thermodynamics. (For example, we could have expressed total energy of disc as $\bigtriangleup { E}= {1\over 2} N k_E T_E $ with out affecting anything.) {\it Hence it can be concluded that the entanglement
entropy  per unit disc area is fixed 
for small discs of radii  $\ell\ll z_I$}. It is 
 also confirmed that the entropy of excitations \eqref{ent34} 
follows the first law relation \cite{Bhattacharya:2012mi, Allahbakhshi:2013rda, Wong:2013gua, Pang:2013lpa, Mishra:2015cpa, Mishra:2016yor, Ghosh:2016fop, Ghosh:2017ygi, Bhattacharya:2019zkb}
\be
\delta s_E=  {1 \over T_E} (\delta \Delta \mathcal{E} + \frac{1}{2}\mu_E\delta \Delta \rho),
\ee
under infinitesimal changes in the bulk quantity, $\delta z_I$.

We summarize  our main observations at first order;
\bea
T_E\propto {1\over \ell^2}, ~~~~
\bigtriangleup s_E = \text{Fixed}, ~~~~
\mu_E\propto r_y T_E, ~~~~
\bigtriangleup {\cal E}\propto N T_E, ~~~~
\Delta \rho= \text{Fixed}, ~~~~
\eea
at a given entanglement temperature.

\subsection{Entanglement entropy of a disc at second order}
\begin{figure}[t]
	\centering
	\includegraphics[scale=0.75]{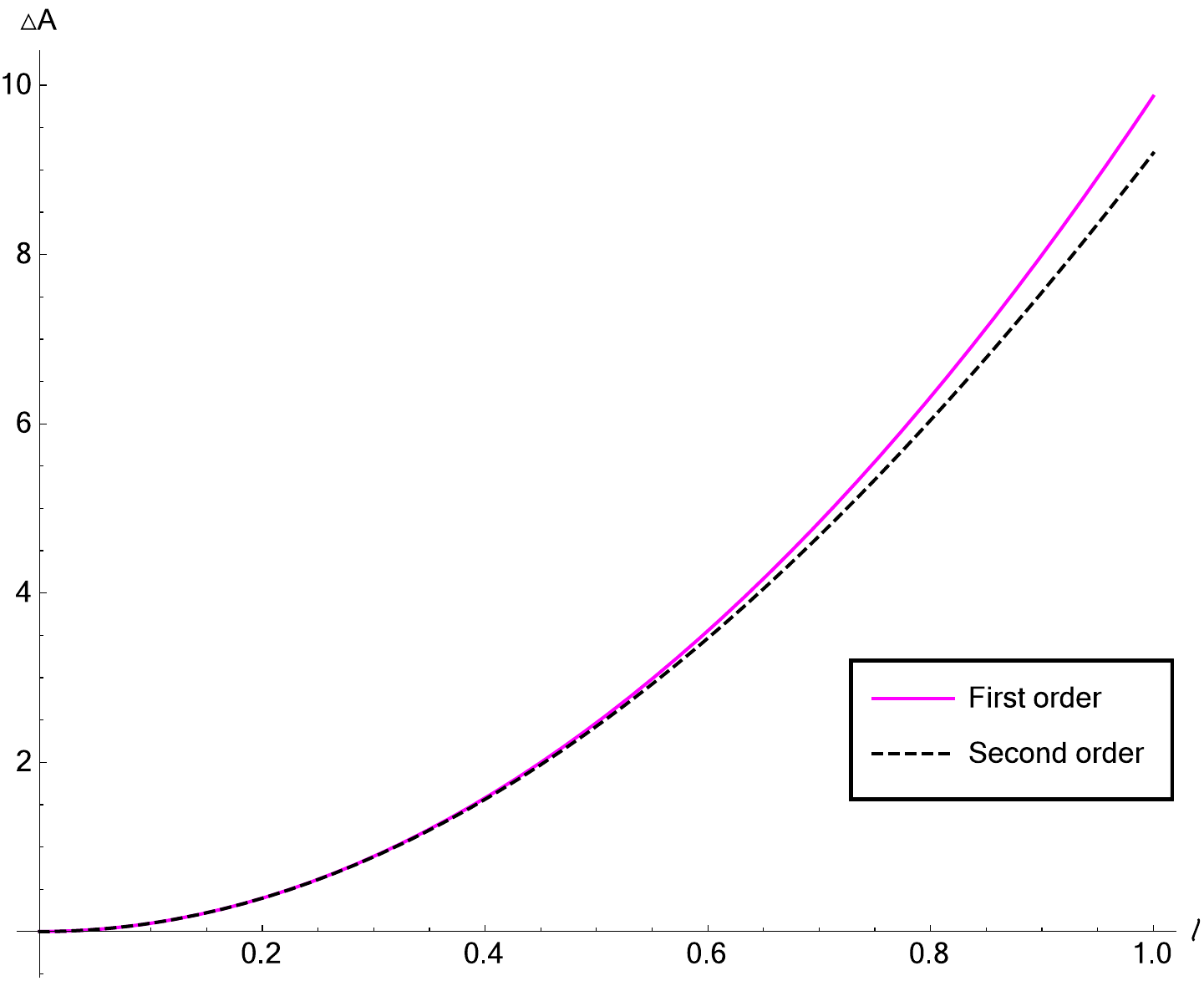}
	\caption{Area difference from $AdS$ ground state for spherical subsystem, the second order correction is negative. Plot drawn by choosing $z_I^2 = 2$ and $L = r_y = q = 1$.}
	\label{spherearea}
\end{figure}

\begin{figure}[h!]
	\begin{subfigure}{0.475\textwidth}
		\centering
		\includegraphics[height=4.75 cm, width=\textwidth]{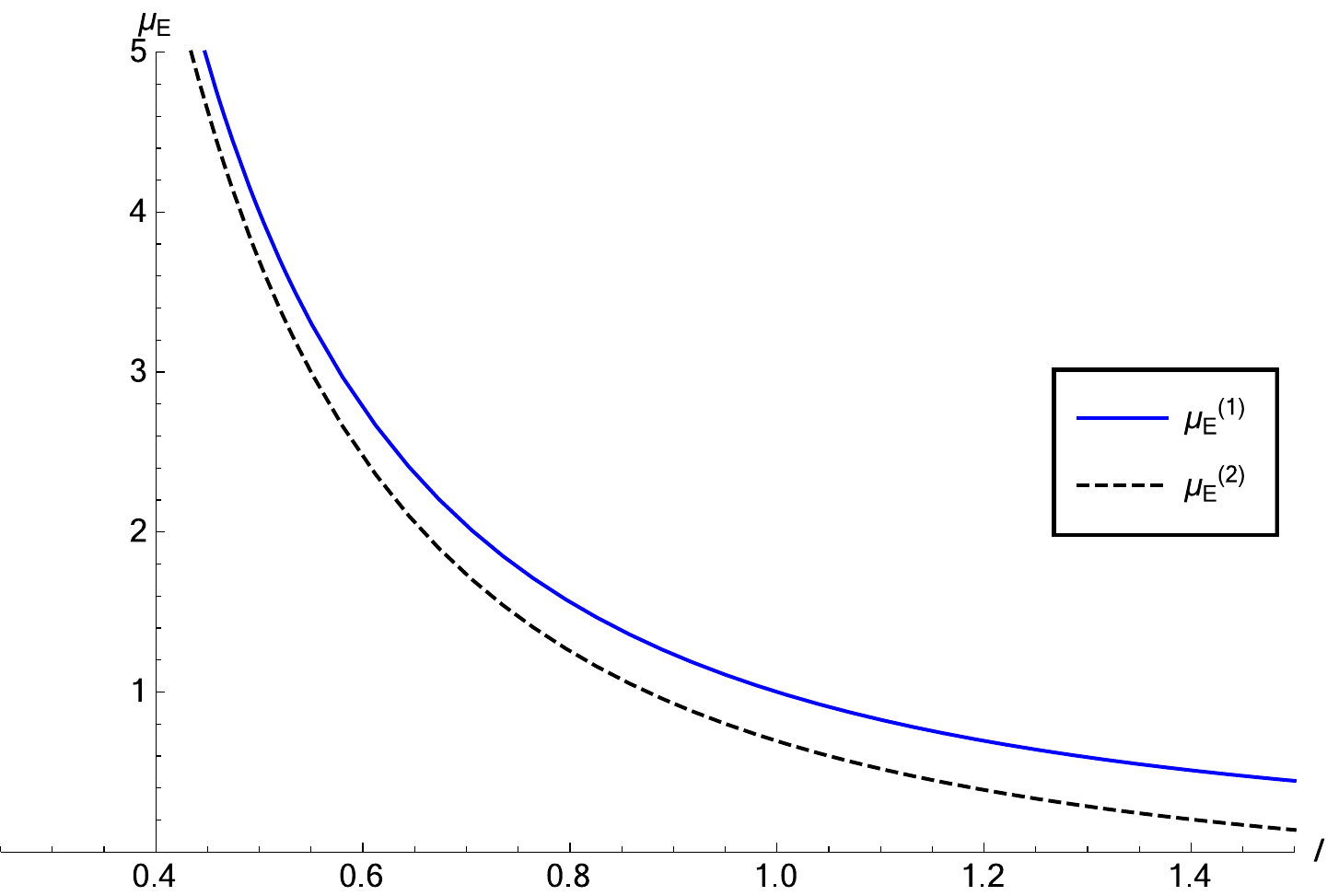}
		\caption{$\mu_E$ vs. $l$}
	\end{subfigure}
	\hfill
	\begin{subfigure}{0.475\textwidth}
		\centering
		\includegraphics[height=4.75 cm, width=\textwidth]{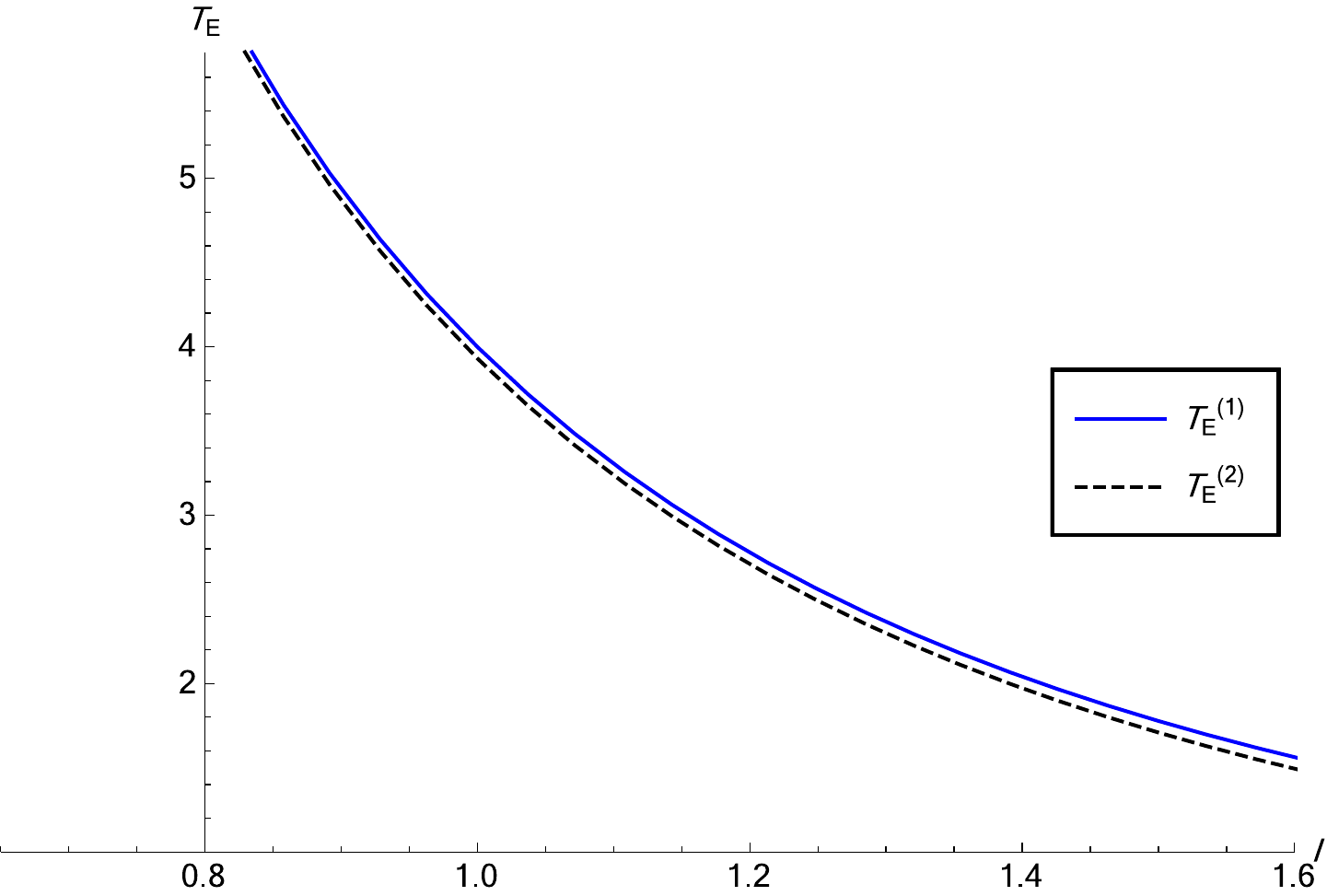}
		\caption{$T_E$ vs. $l$}
	\end{subfigure}
	\caption{The unbroken and dashed curves display the behaviour of the uncorrected and corrected quantities, respectively; both the entanglement temperature and chemical potential decrease due to higher order corrections. The plots were drawn by setting $z_I^2 = 2$ and $L = r_y = q = 1$.}
	\label{fig1}
\end{figure}

Let us now consider corrections to holographic entanglement entropy at next higher order. It is somewhat easier to calculate when one chooses $z(r)$ parameterization, so let us rewrite the integral as
\begin{equation}
	\mathcal{A}_{\gamma} = 2\pi L^2\int_{0}^{1}dr \frac{r\sqrt{1+z'^2}}{z^2}h^{\frac{1}{2}}\,,
\end{equation}
where we rescaled $r$ and $z$ to the dimensionless variables $\frac{r}{\ell}$ and $\frac{z}{\ell}$. It suffices to obtain the embedding up to first order to get the entanglement at second order \cite{Blanco:2013joa, Bhattacharya:2019zkb}. So, we expand $z(r)$ as $z(r) = z_{(0)} + z_{(1)} + \cdots$, where $z_{(0)} = \sqrt{1 - r^2}$ and $z_{(1)}$ satisfies the equation
\begin{equation}\label{eqnsec}
	z_{(1)}'' + \frac{1-2r^2}{r(1-r^2)}z_{(1)}'-\frac{2}{(1-r^2)^2}z_{(1)} = \frac{1}{\sqrt{1-r^2}}\,,
\end{equation}
with the boundary conditions: $z_{(1)}'(0) = 0$ and $z_{(1)}(\ell) = 0$. One can check that a consistent solution to equation \eqref{eqnsec} is
\begin{equation}\label{embdsec}
	z_{(1)} = - \frac{1-r^2-2\sqrt{1-r^2}+2\ln \left(1+\sqrt{1-r^2}\right)}{2\sqrt{1-r^2}}.
\end{equation}
Therefore, the area integral now acquires a new contribution $\mathcal{A}_\gamma = \mathcal{A}_0 + \mathcal{A}_{1} + \mathcal{A}_2$ where
\begin{equation}\label{area2}
	\mathcal{A}_2 = 2\pi L^2\frac{\ell^4}{z_I^4}\left(\frac{5}{8} - \ln 2\right),
\end{equation}
which is negative as expected. The area difference from pure $AdS$ at both orders is plotted in figure \ref{spherearea} .
Total entropy of the disc at this order will be
\begin{equation}\label{discent2}
	S_E^{(2)} = S_E^{(0)} + \frac{\pi L^2}{4 G_4}\frac{\ell^2}{z_I^2}\left(1 + \frac{\ell^2}{z_I^2}\left(\frac{5}{4} - 2\ln 2\right)\right).
\end{equation}
So that the variation of entropy density, at second order, becomes:
\begin{equation}\label{entden2}
	\delta s_E^{(2)} = \frac{L^2}{4 G_4}\left(1 + \frac{\ell^2}{z_I^2}\left(\frac{5}{2} - 4\ln 2\right)\right)\delta\left(z_I^{-2}\right).
\end{equation}
As previous, we wish to express \eqref{entden2} as a `first law' like relationship. We find that one way to achieve this is to absorb all second order corrections to a modified temperature and chemical potential, this method was first used in \cite{Mishra:2015cpa} although they worked with differences rather than variation as we do. To this end, we first note that the turning point $z_*$ should be corrected at $\mathcal{O}\left(\frac{\ell^2}{z_I^2}\right)$ as
\begin{equation*}
	z_* \equiv z(0) = \ell + \frac{\ell^3}{z_I^2}\left(\frac{1}{2} - \ln 2\right).
\end{equation*}
The chemical potential, defined in equation \eqref{chempotdef}, can be expressed including $\mathcal{O}(\frac{\ell^2}{z_I^2})$ corrections as
\begin{align}
	\mu_E^{(1)} \simeq& \frac{L^2r_y}{q\ell^2}\left(1+\frac{\ell^2}{z_I^2}\left(\frac{1}{2} - \ln 2\right)\right)^{-2}\left(1+\frac{\ell^2}{z_I^2}\right)^{-1}\,, \nonumber \\
	=& \frac{L^2r_y}{q\ell^2}\left(1 - \frac{\ell^2}{z_I^2}\left(2 - 2\ln 2\right)\right).
\end{align}
So we get
\begin{align*}
	\mu_E^{(1)}\delta\Delta \rho ~=&~ \frac{2 L^3r_y}{qG_5\ell^2}\left(1 - \frac{\ell^2}{z_I^2}\left(2 - 2\ln 2\right)\right)\delta\left(z_I^{-2}\right),\\
	=&~ \frac{L^2}{\pi G_4\ell^2}\left(1 - \frac{\ell^2}{z_I^2}\left(2 - 2\ln 2\right)\right)\delta\left(z_I^{-2}\right),
\end{align*} 
while the energy remains the same as defined in \eqref{enr34}. From equation \eqref{entden2}, a bit of paperwork then leads to the following result
\begin{equation}\label{spherelaw2}
	\delta s_E^{(2)} = \frac{1}{T_E^{(2)}}\left(\delta\Delta\mathcal{E} + \frac{1}{2}\mu_E^{(1)}\delta\Delta\rho \right),
\end{equation}
where $T_E^{(2)}$ denotes the `entanglement temperature' at second order, which is given by
\begin{align}\label{spheretemp2}
	T_E^{(2)} &= \frac{\delta \Delta \mathcal{E} + \frac{1}{2}\mu_E^{(1)}\delta \Delta \rho}{\delta \Delta s_E^{(2)}},\nonumber \\
	&=\frac{\frac{L^2}{\pi G_4\ell^2}\left[1 - \frac{\ell^2}{z_I^2}\left(1 - \ln 2\right) \right]}{\frac{L^2}{4 G_4}\left[1 - \frac{\ell^2}{z_I^2}\left(4\ln 2 - \frac{5}{2}\right) \right]},\nonumber \\
	&\simeq T_E^{(1)}\left[1 + \frac{\ell^2}{z_I^2}\left(5\ln 2 - \frac{7}{2}\right) \right],
\end{align}
where $T_E^{(1)}$ stands for the first order temperature, defined in eqn. \eqref{spheretemp1}. The term in parentheses is a negative number, so second order correction to `entanglement temperature' results in its sharper fall. See figure \ref{fig1} for an illustration of this behaviour.

\par Some comments are in order to justify equation \eqref{spherelaw2}, we have seen that for small enough subsystem size $(\ell \ll z_I)$, the change in entanglement entropy at first order in our perturbative calculation follows a relationship akin to the first law of thermodynamics. If one considers this relationship an actual `law' for entanglement entropy, one must find a consistent way to describe new contributions at higher orders. Equation \eqref{spheretemp2} proposes that at second order, the chemical potential as well as the entanglement temperature should be corrected to keep the law intact. In fact, we expect this procedure to work at all higher orders. It could be thought that a more accurate measure of these quantities are obtained as one climbs the perturbation ladder.
\section{ Entanglement entropy of narrow  strip}\label{sec4}
 
We now consider a strip like subsystem with  coordinate
width $-\ell/2\le x_1\le \ell/2$, and the  range of $x_2\in[0,l_2]$, such that
$l_2\gg \ell$. The straight line  boundary of the two-dimensional strip
is identified with the boundary  of the  RT surface in the bulk 
at constant time. 
The area functional of this static  surface is 
\bea\label{str10}
{\cal A}_\gamma=2 L^2 l_2  \int^{z_\ast}_{\epsilon} dz 
{\sqrt{1+x_1'^2}\over z^2} h^{1\over2}\,. 
\eea
For small width  $\ell\ll z_I$, we 
make a perturbative expansion of the integrand. 
The extremal surface satisfies the following equation
\bea 
x_1'={z^2\over z_\ast^2} {1\over \sqrt{{h\over h_\ast}-{z^4\over z_\ast^4}}}\,,
\eea
where $h_\ast\equiv h(z_\ast)$.
We have  specific boundary conditions such that
near  the  spacetime boundary 
$x_{1}|_{z=0}=\ell/2$ and  
the turning point is given by $x_{1}|_{z\sim z_\ast}=0$. 
This leads to the first integral of the following type
\bea\label{str11} 
\ell=2\int^{z_\ast}_0 dz{z^2\over z_\ast^2} {1\over \sqrt{{h\over h_\ast}-{z^4\over z_\ast^4}}}\,,
\eea
which gives rise to a perturbative expansion in ${z_\ast \over z_I}$ 
\bea \label{str12}
\ell=z_\ast
\left(b_0 +{z_\ast^2 \over2 z_I^2} I_1 + \cdots\right)\,,
\eea
where coefficients are expressible as Beta-functions $b_0=
{1\over 4} B\left(\frac{3}{4}, \frac{1}{2}\right)$ and $I_1= 
{1\over 4}(B(\frac{3}{4},-\frac{1}{2}) -B(\frac{5}{4},-\frac{1}{2}))$.
The equation \eqn{str12} can be inverted and expressed as a perturbative expansion of the turning point
\bea z_\ast=z_\ast^{(0)} \left(1 -{z_\ast^{(0)2} \over z_I^2} {I_1\over2 b_0} + \cdots \right),
\eea
where $z_\ast^{(0)}\equiv {\ell \over 2b_0}$ is the turning point in the 
absence of excitations.

The leading area of strip can be evaluated using the tree level values
\bea
{\cal A}_0 &&= 
2 L^2 l_2 \int^{z_\ast^{(0)}}_{\epsilon} dz 
{\sqrt{1+x_{1(0)}'^2}\over z^2}\,,\br
&&={2 L^2 l_2\over z_\ast^{(0)}} \int^{1}_{\epsilon \over z_\ast^{(0)}} 
d\zeta {1\over\zeta^2 \sqrt{1-\zeta^4}}\,,
\br &&
=2 L^2 l_2 \left({1\over\epsilon}-{2(b_0)^2\over \ell}\right).
\eea
while the first order contribution is evaluated as
\bea\label{foa}
{\cal A}_1 &&= 
2 L^2 l_2 \int^{z_\ast}_{0} dz  
{\sqrt{1+x_{1(0)}'^2}\over 2 z_I^2}\,,\br
&& = L^2 l_2\left({a_1 z_\ast^{(0)} \over z_I^2 }\right).
\eea
where the coefficient $a_1={1\over 4} B\left(\frac{1}{4},\frac{1}{2}\right)$.
The entanglement entropy of small strip  up to first order is then given by
\bea
S_E^{strip} = \frac{\mathcal{A}_0 + \mathcal{A}_1}{4G_5} = {L^2 l_2 \over 2 G_{4}}  
\left({1\over \epsilon} -{2b_0^2\over \ell}
+{a_1\over 4b_0}~\frac{\ell}{z_I^2} \right).
\eea

Now any small  change in the bulk parameter 
($\delta z_I$) will necessarily effect the entanglement entropy at first order.For a fixed width $\ell$, we find the change in entropy per unit area of the strip as
\bea\label{jh4}
\delta s_E^{strip} \equiv {\delta S_E^{strip}\over l_2 \ell}
=  {L^2 \over 8 G_{4}} {a_1\over b_0}
\delta \left( {z_I^{-2}} \right),
\eea
which is complete expression up to first order. Once again we find that the right hand side is independent of $\ell$, as it was also in the case of a disc. Following from the  disc case in the previous section, the effective chemical potential for strip subregion is
\bea
\mu_E=
{L^2 r_y\over q z_\ast^2}\simeq 
{4 b_0^2 L^2 r_y\over q \ell^2}\,.
\eea
From here and \eqref{rho12}, let us define for the strip
\bea\label{enr35} 
\bigtriangleup {\cal E} \equiv \frac{1}{2}
\mu_E . \bigtriangleup \rho = {4 L^3r_y \over  G_5\,q}\frac{b_0^2}{z_I^2 \ell^2 } = \frac{2 L^2}{\pi G_4}\frac{b_0^2}{z_I^2\ell^2}.
\eea
This is like the disc result in \eqn{enr34}, $i.e.$ 
$\bigtriangleup {\cal E}\propto T_E$.
Using  \eqn{enr35} we conclude  that the entanglement
entropy density \eqn{jh4} of the strip subsystems 
also conforms to the first law relation
\bea \label{flfo}
\delta s_E={1\over T_E} \left(\delta \Delta \mathcal{E} + \frac{1}{2}\mu_E\delta \Delta \rho\right). \eea
where for the strip, entanglement temperature is
defined as $T_E={8 b_0^3 \over a_1}{4\over \pi \ell^2}$ 
in  3-dimensional  Lifshitz theory.

\subsection{Strip entropy at second order}

It is instructive to find out the change in entanglement 
entropy at higher orders in $\frac{\ell^2}{z_I^2}$ and 
interpret its thermodynamic property, here we include the 
results at $\mathcal{O}\left(\frac{\ell^4}{z_I^4}\right)$. \\ 
\par The turning point $z_*$, as discussed before in \eqn{str11} and \eqn{str12}, could be related to the strip-width $\ell$ as
	\begin{equation}\label{turnpt2}
		z_* = 
\frac{z_*^{(0)}}{1 + \frac{z_*^{(0) 2}}{2z_I^2} \frac{I_1}{b_0} - 
\frac{z_*^{(0) 4}}{8z_I^4} \left(\frac{I_2}{b_0} + \frac{4I_1^2}{b_0^2}\right)}\,,
	\end{equation}
	where the new co-efficient $I_2$ can be expressed as: 
$I_2 = \frac{1}{8}\big(2B(\frac{3}{4}, -\frac{3}{2}) - 
3B(\frac{5}{4}, -\frac{3}{2})\big)$. With the help of 
\eqref{turnpt2}, the area integral \eqn{str10} now reads
 		$\mathcal{A}_{\gamma} = \mathcal{A}_0 + \mathcal{A}_1 + \mathcal{A}_2$,
 where $\mathcal{A}_0$ and 
$\mathcal{A}_1$ are as obtained before.
  The second order contribution is
	\begin{equation}
\mathcal{A}_2=
- \frac{2 L^2 l_2}{z_*^{(0)}} \frac{z_*^{(0)4}}{8z_I^4}\left(\frac{4a_0I_1^2}{b_0^2} + \frac{2I_1J_1}{b_0}\right).
	\end{equation}
	The new coefficients introduced in above expression are listed below
	\begin{align*}
		a_0 &= -\frac{1}{4}B\left(\frac{3}{4}, \frac{1}{2}\right) = -b_0\,, \\
		J_1 &=\frac{1}{4}\left(B\left(\frac{3}{4}, -\frac{1}{2}\right) 
+ 3B\left(\frac{1}{4}, -\frac{1}{2}\right) \right).
	\end{align*}

\begin{figure}[t]
		\centering
		\includegraphics[scale=0.75]{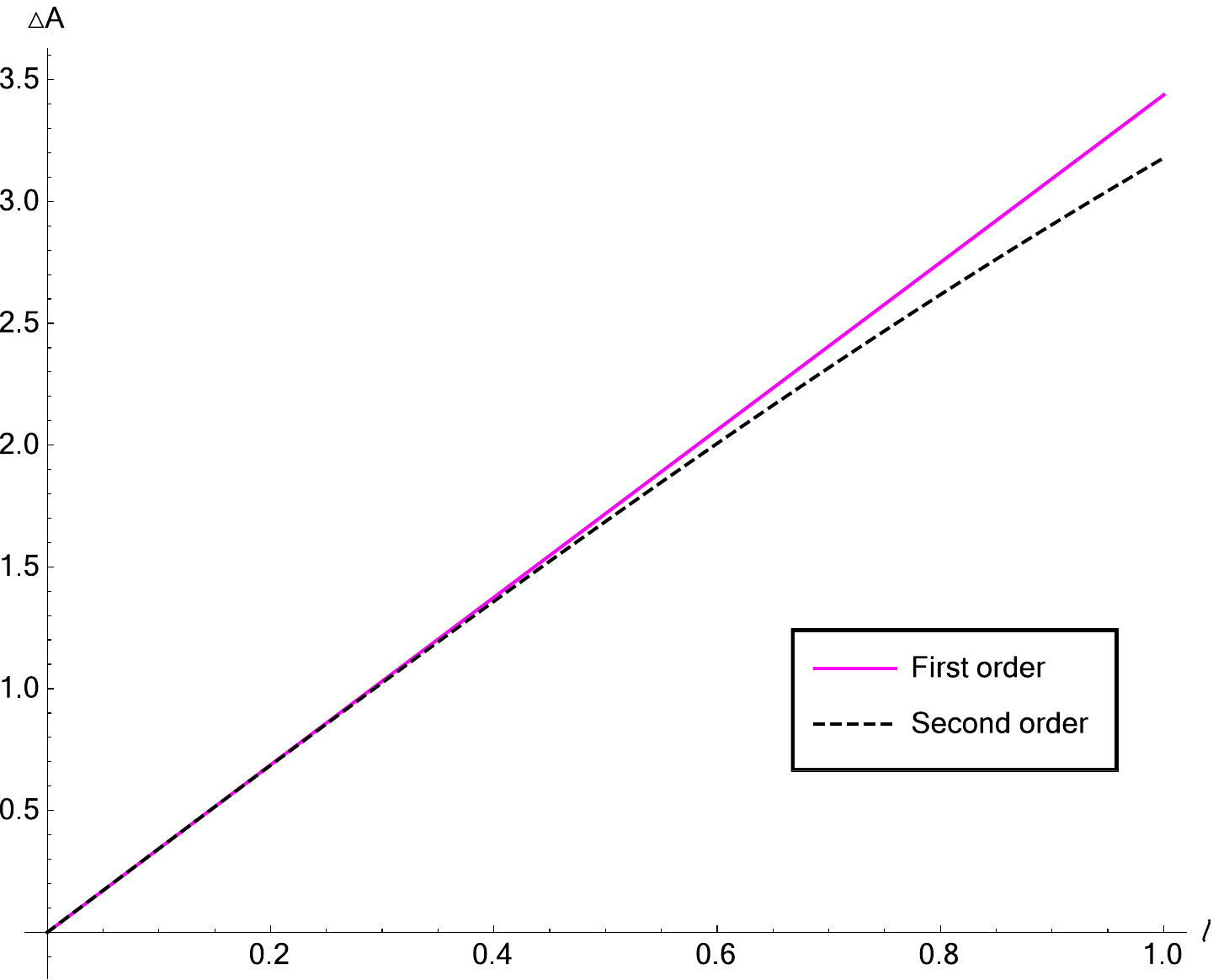}
		\caption{The area difference at first and second order of perturbation analysis for strip subsystem, plots drawn by choosing $z_I^2 = 2$ and $L=r_y=q=l_2=1$.}
		\label{striparea}
	\end{figure}

\begin{figure}
	\begin{subfigure}{0.475\textwidth}
		\centering
		\includegraphics[height=4.75 cm, width=\textwidth]{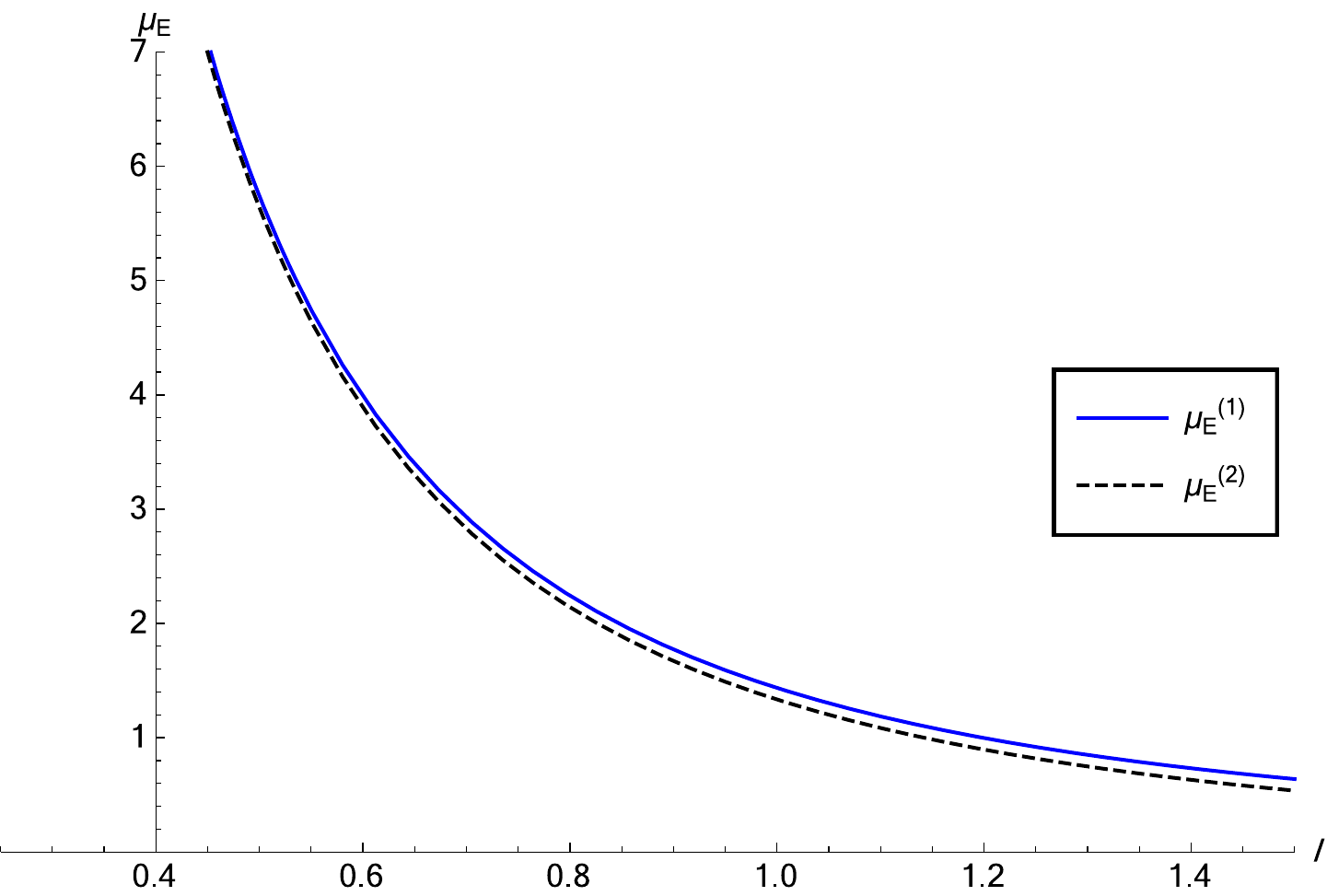}
		\caption{$\mu_E$ vs. $\ell$}
	\end{subfigure}
	\hfill
	\begin{subfigure}{0.475\textwidth}
		\centering
		\includegraphics[height=4.75 cm, width=\textwidth]{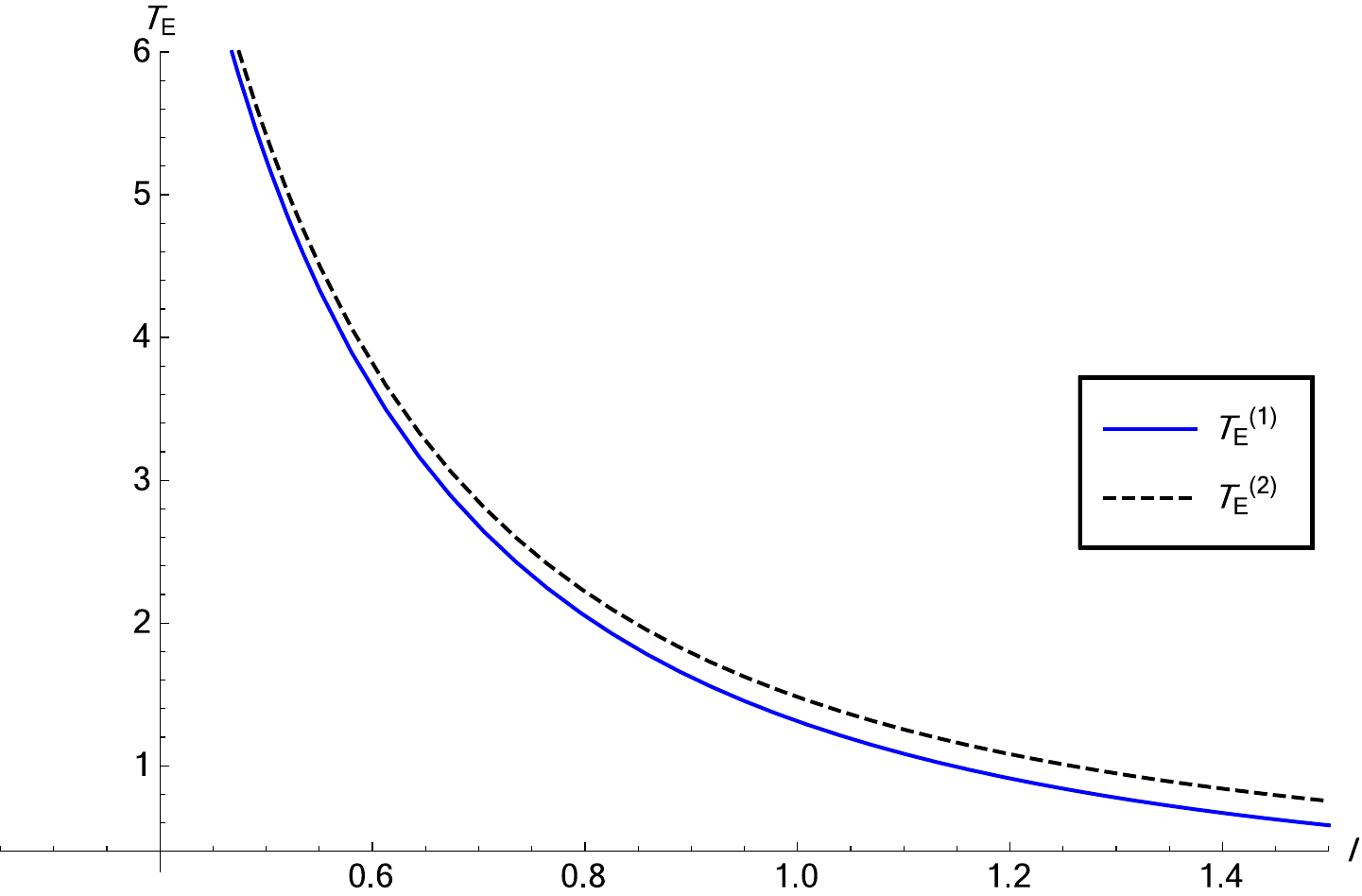}
		\caption{$T_E$ vs. $\ell$}
	\end{subfigure}
	\caption{The unbroken and dashed curves display the behaviour of the uncorrected and corrected quantities, respectively; the entanglement temperature is found to increase due to higher order corrections while the chemical potential decreases.The plots were drawn by setting $z_I = 2$ and $L = r_y = q = G_5 = 1$.}
	\label{fig2}
\end{figure}

After some simplification the contribution to the area of the RT 
surface at second order turns out to be
	\begin{equation}
		\mathcal{A}_2 = -\frac{L^2 l_2\ell}{64} \frac{\ell^2}{z_I^4}\frac{1}{b_0^2}\left(\frac{a_1^2}{b_0^2} - 1 \right).
	\end{equation}
	The coefficient $a_1$ has already been defined in eq. \eqref{foa}. Hence, the total entanglement entropy density, at second order in perturbation theory, becomes
	\begin{equation}\label{kl23}
		s_E^{(2)} = s_E^{(0)} + \frac{L^2}{8 G_4}\frac{1}{z_I^2}
\frac{a_1}{b_0}\left(1 - \frac{\ell^2}{z_I^2} \frac{1}{32b_0^2}\left(\frac{a_1^2}{b_0^2} - 1\right)\right).
	\end{equation}
	The area difference including second order correction has been shown in figure \ref{striparea}.
	To write down the `first law' we need to rewrite the expression for $s_E^{(2)}$ in terms of variation in $\mathcal{E}$ and $\mu_E \Delta \rho$; recall that the chemical potential was defined as the value of the gauge potential at the turning point. Here, it is sufficient to compute $\mu_E$ up to first order
	\begin{equation*}
		\mu_E^{(1)} \simeq \frac{L^2}{z_*^2}\left(1 - \frac{z_*^2}{z_I^2} \right) = \frac{L^2r_y}{qz_*^{(0)2}}
\left(1 + \frac{z_*^{(0)2}}{z_I^2}(\frac{I_1}{b_0^2} - 1) \right).
	\end{equation*}
	So that,
	\begin{align*}
		\mu_E^{(1)}\delta \Delta \rho ~=&~ \frac{L^3r_y}{qG_5}\frac{8b_0^2}{\ell^2}\left[1 + \frac{\ell^2}{z_I^2}\frac{1}{8b_0^2}\left(\frac{a_1}{b_0} - 3\right) \right]\delta\left(z_I^{-2}\right),\\
		=&~ \frac{L^2}{2\pi G_4}\frac{8b_0^2}{\ell^2}\left[1 + \frac{\ell^2}{z_I^2}\frac{1}{8b_0^2}\left(\frac{a_1}{b_0} - 3\right) \right]\delta\left(z_I^{-2}\right).
	\end{align*}
A little effort, then, allows us to write
	\begin{equation}\label{law2}
		\delta s_E^{(2)} = \frac{1}{T_E^{(2)}}\left(\delta \Delta \mathcal{E} + \frac{1}{2}\mu_E^{(1)}\delta \Delta \rho\right).
	\end{equation}
	Here, $T_E^{(2)}$ stands for the modified entanglement temperature at second order.
	\begin{align}\label{temp2}
		T_E^{(2)} &= \frac{\delta \Delta \mathcal{E} + \frac{1}{2}\mu_E^{(1)}\delta \Delta \rho}{\delta \Delta s_E^{(2)}},\nonumber \\
				  &= \frac{4}{\pi \ell^2}\frac{8b_0^3}{a_1}
					\left[1 + \frac{\ell^2}{z_I^2}\frac{1}{16b_0^2} \left(\left(\frac{a_1}{b_0} - 3 \right) + \left(\frac{a_1^2}{b_0^2} - 1\right) \right) \right] \nonumber \\
				  &= T_E^{(1)}\left[1 + \frac{\ell^2}{z_I^2}\frac{1}{16b_0^2}\left(\left(\frac{a_1}{b_0} - 1\right)\left(\frac{a_1}{b_0} + 2 \right) - 2 \right) \right]
	\end{align}
Where by $T_E^{(1)}$, we refer to the temperature at first order defined in equation \eqref{flfo}, the numerical value of $\frac{a_1}{b_0} \approx 2.188$, so the correction at this order results in an increase of $T_E$, albeit by a tiny amount. The uncorrected and corrected temperatures are plotted in figure \ref{fig2}.

\subsection{Numerical results for strip subsystem}
\begin{figure}[t]
	\centering
	\includegraphics[scale=0.5]{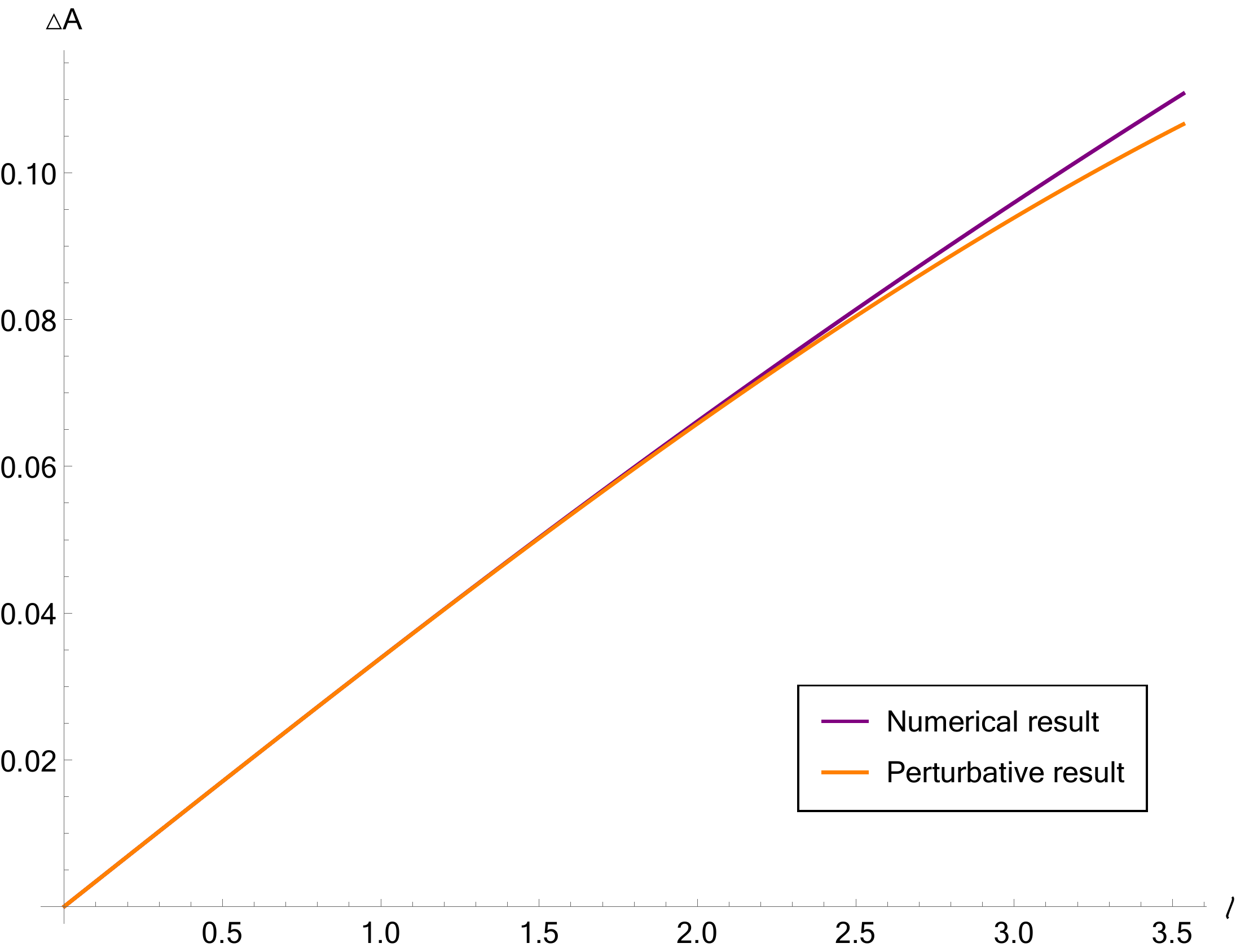}
	\caption{Numerical plot of area difference from $AdS$ ground state for strip subsystem and comparison with second order perturbation series analysis. The pre-factor in \eqref{str10} was ignored in the plot.}
	\label{stripnumeric}
\end{figure}

We end this section with a comparison of our perturbative results with some numerical analysis. For the numerical computation we chose $z_I = 4$ and used \eqref{str11} to obtain corresponding lengths $\ell$ of the sub-region for different choices of the turning point $z_{*}$. We also obtain the area difference $\Delta \cal{A}$ from \eqref{str10} for the same $z_{*}$ values and plot the two sets against each other. The output is summarized in figure \ref{stripnumeric}. From the graph we conclude that a second order perturbation series analysis is trustworthy for small strip-width.

\section{Conclusion}\label{sec5}

The Lifshitz background  
$Lif_4^{(2)}\times {S}^1\times S^5$ of the massive type IIA theory
allows exact  excitations which couple to massless modes of string in the IR. We calculated the entanglement entropy of the theory
at the boundary of these spacetimes, both for strip as well as disc shaped systems. At leading order, we found that the entropy density of the excitations remains fixed and does not grow with $\ell$, the subsystem size, so long as $\ell\ll z_I$. We find that this behaviour is consistent with the fact that energy density of the excitations itself behaves as  $\bigtriangleup{\cal E} \propto 1/\ell^2$, which is in agreement with $\bigtriangleup{\cal E}
\simeq \frac{1}{2}\mu_E \bigtriangleup\rho $. Note that the entanglement temperature itself goes as $T_E\propto{1\over \ell^2}$.

But this entanglement behaviour is quite different in comparison to the relativistic CFTs, where the entropy density of excitations grows linearly with the subsystem size, while the energy density of excitations remains fixed. Nevertheless we have found that the first law of entanglement thermodynamics
\begin{equation}
	\delta s_E = \frac{1}{T_E}\left(\delta \Delta \mathcal{E} + \frac{1}{2}\mu_E\delta \Delta \rho\right),
\end{equation} 
holds good if we accept the hypothesis that the energy of a subsystem in the Lifshitz background \eqref{sol2a9} is given by
$$\bigtriangleup{ E}\simeq \mu_E N 
\simeq {1\over 2} N k_E T_E\,. $$
Our results appear to indicate an 
equipartition nature of the entanglement thermodynamics for non-relativistic Lifshitz
system. But this is perhaps true only for 
the high entanglement temperature regime (i.e. small $\ell\ll z_I$).
\par Further, we studied what happens to the first law of entanglement if we assume it to remain valid beyond the leading order. There is lack of consensus on this aspect, despite there being enough evidence for it to be a natural feature at first order. We discussed how the first law could be extended up to second order by making use of appropriately modified chemical potential and entanglement temperature. We think this is necessary because otherwise, we need to look for a new quantity at each higher order to account for the corrections; while the entanglement entropy, like its thermal counterpart should depend only on the energy and conserved charges of the theory. Such redefinition should work at all orders, thereby allowing the `first law of entanglement thermodynamics' to be obeyed quite generally, irrespective of the degree of perturbation theory.
\par It would be interesting to obtain the HEE numerically for ball subsystems and compare with our perturbative results. This, however, involves solving boundary value problem and proves to be non-trivial. Another interesting problem is to consider shape dependence of holographic entanglement entropy in similar spirit to \cite{Fonda:2015nma, Cavini:2019wyb}. We hope to return to these problems in future.
\vskip 0.5cm
\begin{acknowledgements} 
HS is thankful to the organisers of the 
AdS/CFT@20 workshop at ICTS Bangaluru 
and the STRINGS-2019 at Brussels for the exciting meetings and warm hospitalities. SM would like to thank Aranya Bhattacharya for useful discussions and help with Mathematica.
\end{acknowledgements} 

\appendix
\section{The winding string charge in massive Lifshitz vacua}

Here we would like to know  the winding number of the string
excitations. The circle compactification of the background \eqn{sol2a9} 
along $y$ direction
gives rise to following 9-dimensional fields (we  set $g_0=1,~\alpha'=1$)
\bea\label{sol2a9a}
&&ds^2_{D=9}= L^2\left(- {dt^2\over  z^4h} +{dx_1^2+dx_2^2\over z^2}+{dz^2\over
z^2}  + d\Omega_5^2 \right) ,\br
&&e^{2\bar\phi}= {1\over h \sqrt{G_{yy}}},
~~~~~A_t=  { L^2\over q  z^2}h^{-1}  \ ,
\eea 
where  $ G_{yy}={L^2\over q^2 h},~h(z)= 1+{z^2\over z_{I}^2}$. The $\bar\phi$
is 9-dimensional dilaton field. 
The corresponding gauge field strength $F_{(2)}=d A$ gives rise to the winding
charge
\bea\label{hj45}
Q&&={  \pi r_y\over G_{10}}\int e^{-{4\bar\phi\over 7}} G^{yy} (\ast_{9} F_{(2)}) \br
&&= {\pi L^6\omega_5  r_y\over G_{10} }\int dx_1 dx_2 
\left({2\over z^2} +{4\over z_I^2}\right) \br
&&= {\pi L  r_y V_2\over G_{5} } \left({2\over z^2} +{4\over z_I^2}\right) \br
&&\equiv Q_{ground-state}+\bigtriangleup Q 
\eea 
where $\omega_5$ is the size of unit 5-sphere. 
The total  charge $Q$, of course, depends on scale $z$, 
 because we are in  asymptotically (non-flat)
Lifshitz spacetime. However,  the contribution purely due to string
excitations is given by $\bigtriangleup Q$. The second term in \eqn{hj45} 
is not affected by $z$ and remains  constant. 
Therefore the net contribution of string excitations 
is 
\bea
\bigtriangleup Q= Q-Q_{ground-state}=
{2\pi L  r_y V_2\over G_{5} } \left({2\over z_I^2}\right) \simeq Q|_{z=\infty}.
\eea 
Alternatively the charge due to string excitations can also be measured near
$z\sim\infty$, where the massive mode gets completely 
 decoupled and only massless
strings survive which contribute to the charge. Net winding number of these
 strings is quantized 
in the units $ N={\bigtriangleup Q \over r_y}$, where $N$ is an integer.
%
	%


\end{document}